\documentclass[amsmath, amssymb, english, aps, prl, twocolumn, reprint,floatfix, superscriptaddress,showpacs]{revtex4-1}

\usepackage[english]{babel}
\usepackage{graphicx}
\usepackage{verbatim} 

\def\tg{t${}_{2g}$}
\begin{document}
\newcommand{\eg}{e${}_{g}$}
\newcommand{\note}[1]{{\bf[[NOTE: #1 ]]}}

\newcommand{\MA}[1]{{\em #1 }}

\author{Jernej~Mravlje}
\affiliation{Coll\`ege de France, 11 place Marcelin Berthelot, 75005 Paris, France}
\affiliation{Centre de Physique Th\'eorique, \'Ecole Polytechnique,
  CNRS, 91128 Palaiseau Cedex, France}
\affiliation{Jo\v{z}ef Stefan Institute, Jamova~39, Ljubljana, Slovenia}
\author{Markus~Aichhorn}
\affiliation{Institute of Theoretical and Computational Physics, TU Graz, Petersgasse 16, Graz, Austria}
\author{Antoine~Georges}
\affiliation{Centre de Physique Th\'eorique, \'Ecole Polytechnique,
  CNRS, 91128 Palaiseau Cedex, France}
\affiliation{Coll\`ege de France, 11 place Marcelin Berthelot, 75005 Paris, France}
\affiliation{DPMC, Universit\'e de Gen\`eve, 24 quai Ernest Ansermet, CH-1211 Gen\`eve, Suisse}

\begin{abstract}
We investigate the origin of the high N\'eel temperature recently
found in $\mathrm{Tc}^{4+}$ perovskites. The electronic structure in
the magnetic state of SrTcO$_3$ and its $3d$ analogue SrMnO$_3$ is
calculated within a framework combining band-structure and many-body
methods.  In agreement with experiment, for SrTcO$_3$ a smaller magnetic
moment and 4 times larger N\'eel temperature are found.  We show
that this is because the Tc compound lies on the verge of the
itinerant-to-localized transition, while the Mn compound lies deeper
into the localized side. For SrTcO$_3$ we predict that the N\'eel
temperature depends weakly on applied pressure, in clear violation of
Bloch's rule, signaling the complete breakdown of the localized
picture.
\end{abstract}

\title{Origin of the high N\'eel temperature in SrTcO$_3$}

%71.27.+a   Strongly correlated electron systems; heavy fermions
%73.23.-b   Electronic transport in mesoscopic systems
%73.23.Hk   Coulomb blockade; single-electron tunneling
%72.15.Qm   Scattering mechanisms and Kondo effect (see also
%               75.20.Hr local moments in compounds and alloys; Kondo effect, valence
%               fluctuations, heavy fermions in magnetic properties and materials)
%73.22.-f   Electronic structure of nanoscale materials: clusters,
%               nanoparticles, nanotubes, and nanocrystals

%75.50.Ee 	Antiferromagnetics
% 75.47.Lx 	Magnetic oxides
% 75.40.Mg 	Numerical simulation studies 

\maketitle 

Recently, antiferromagnetism persisting to very high temperatures
exceeding 1000K has been reported for SrTcO$_3$ \cite{rodriguez11,
  thorogood11} and other $\mathrm{Tc}^{4+}$ perovskites
\cite{avdeev11}.  Because Tc is a radioactive element, those
perovskites have been scarcely investigated, and for many of them only
structural properties are known. This discovery is especially striking
since robust magnetism with high transition temperatures has been
reported up to now only for 3d transition metals and their oxides,
which realize large moments in localized 3d shells.  Indeed, magnetism
in 4d materials with more extended orbitals has been rarely
found. Among the few reported cases, the antiferromagnetic
Ca$_2$RuO$_4$ has a low N\'eel temperature of 110K \cite{nakatsuji97}, and
SrRuO$_3$ is an itinerant ferromagnet with $T_c=160 K$ and a small
moment 1.5$\mu_B$ \cite{kanbayasi76,cao97}.

The N\'eel temperature of SrTcO$_3$ appears remarkably high especially
when compared to $T_N=260K$ \cite{takeda74} found in its 3d analogue
SrMnO$_3$. The comparison is in order, as both compounds have the same
$G$- type antiferromagnetic order and adopt a simple cubic structure
in the relevant temperature range. Remarkably, the Tc magnetic moment
found is smaller, 2.1$\mu_B$\cite{rodriguez11}, compared to 2.6$\mu_B$
\cite{takeda74} for the Mn- compound. In Ref.~\cite{rodriguez11}, the
small Tc moment is attributed to covalency \cite{hubbard65}, that is
to the moment residing in orbitals with considerable weight on oxygen
sites, where it overlaps with the opposite moment of the neighboring
Tc atom. However, the covalency is large in the electronic structure
of SrMnO$_3$ as well~\cite{[{The covalency occurs as the extended d
      orbitals hybridize strongly with oxygens, with strength of
      hybridization $t_{pd}^2/\Delta$ with $t_{pd}$ the overlap and
      $\Delta$ the separation of oxygens from the $t_{2g}$
      states. While $t_{pd}$ is larger in SrTcO$_3$ due to the larger
      spatial extension of 4d orbitals, the larger $\Delta$
      compensates for this and the covalency of the two compounds is
      found to be comparable. See also: }]sondena06} and thus cannot
explain the smaller moment or the much higher N\'eel temperature
\cite{rodriguez11} of the Tc compound.

In this Letter we resolve this puzzling situation by calculating the
electronic structure of SrTcO$_3$ and SrMnO$_3$ in a theoretical
framework which combines band-structure and many-body methods, as
appropriate for $d$ shells with strong correlations.  Our approach
allows us to put forward qualitative explanations, which we support by
quantitative calculations.  First, we point out that Tc$^{4+}$
compounds are unique among 4d oxides: Because they have a half-filled
$t_{2g}$ shell, the moderate Coulomb repulsion associated with
extended 4d orbitals is nevertheless sufficient to localize the
electrons.  Second, we show that the key property of SrTcO$_3$ is that
it is located close to the metal-insulator transition of the
paramagnetic state and that the N\'eel temperature is maximal there,
as anticipated in Ref.~\cite{demedici11}. In contrast, SrMnO$_3$ is
deeper into the insulating side. Third, we relate the smaller magnetic
moment in SrTcO$_3$ to the corresponding larger charge fluctuations in
this compound. The effects of covalency with oxygen are carefully
analyzed as well.

The authors of Ref.~\onlinecite{rodriguez11} recognized the more itinerant nature of
SrTcO$_3$ but still based their explanation of magnetism on a large Heisenberg
exchange interaction $J_H$, a point of view forcefully advocated in
Ref.~\onlinecite{franchini11}.  Under pressure, $J_H$ increases and most
oxides obey Bloch's rule \cite{bloch66}, $\alpha=d \log(T_N)/d
\log V \approx -3.3$.  We demonstrate that, for SrTcO$_3$, a small
value of $\alpha$ is found, hence violating Bloch's rule. This is
a hallmark of SrTcO$_3$ being at the verge of an
itinerant-to-localized transition and signals the inapplicability of
the localized description.

\paragraph{ Methods.} 
We use the theoretical framework which combines dynamical mean-field
theory (DMFT) and density-functional theory in the local density
approximation (LDA), in the charge self-consistent implementation of
Refs.\cite{aichhorn09,aichhorn11} based on the Wien2K package
\cite{wien2k}.  The cubic crystal structure and G-type
antiferromagnetic unit cell have been used for both compounds.  Wannier
\tg\ orbitals $\psi_m$ are constructed out of Kohn-Sham bands within
the energy window containing $t_{2g}$ bands, that is within
$[-2.5,1.2]$ and $[-2.0,0.8]\,$eV for Tc and Mn compounds,
respectively. We use a fully rotationally invariant interaction in the
form $H_I =(U-3J) n(n-1)/2 -2J S^2 -1/2 J T^2$, with $U$ the on-site
Hubbard interaction, $J$ Hund's rule coupling, and $n$, $S$, and $T$ the
total charge, spin and orbital momentum on the atom, respectively.  We
solve the DMFT quantum impurity problem by using the TRIQS
toolkit\cite{TRIQS} and its implementation of the continuous-time
quantum Monte-Carlo algorithm \cite{gull11,Legendre}.  The interaction
parameters were chosen as $U=2.3$eV, $J=0.3$eV for Tc and $U=3.5$eV,
$J=0.6$eV for Mn compound. For the former, we set the parameters
close to the values of the Ru compounds \cite{mravlje11}. For the
latter, we checked that the position of the lower Hubbard band agrees
reasonably well with photo emission spectroscopy \cite{kang08}. The
results do not change significantly if the interaction parameters are
varied in the physically plausible range.  For SrMnO$_3$ we checked
also that the \eg\ degrees of freedom are inactive by performing a
5-orbital calculation including \eg\ orbitals.  The magnetization
found then is indistinguishable from the results reported below using
only t$_{2g}$ states.  To account for the covalency effects, the
magnetic moment is recalculated also with a different choice of
Wannier orbitals $\Psi_m$ constructed from the larger energy window
including all the oxygen bands.

\begin{figure}
 \begin{center}
   \includegraphics[width=0.95\columnwidth,keepaspectratio]{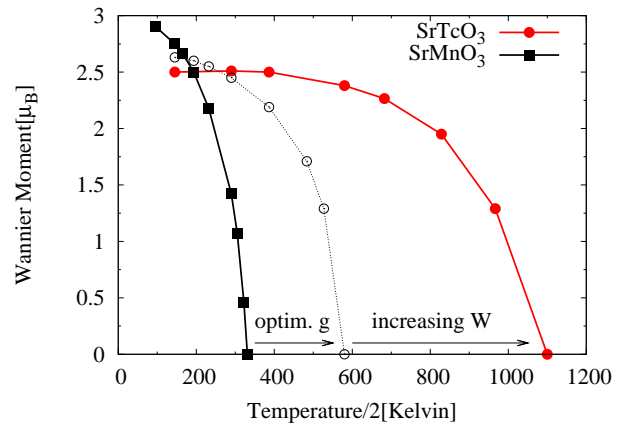}
   \end{center}
   \caption{\label{fig:moment} Solid lines: Temperature dependence of
     SrTcO$_3$ (filled circles) and SrMnO$_3$ (squares) `Wannier'
     magnetic moments in orbitals $\psi_m$ (see discussion in the
     text). Note that the $x$ axis has been rescaled due to systematic
     overestimation of $T_N$ in DMFT \cite{lichtenstein01}. Open
     circles: Results obtained for a hypothetical compound 
     having the SrMnO$_3$ band structure but artificially reduced
     interactions yielding a maximal N\'eel temperature.}
 \end{figure}

\paragraph{Results} 
Our main result accounting for the key qualitative aspects reported in
experiments is presented in Fig~\ref{fig:moment}.  Because all spatial
fluctuations are neglected in DMFT, which is a mean-field approach,
the absolute magnitude of the calculated N\'eel temperature exceeds
the real values by about a factor of $2$, as documented in previous
work\cite{lichtenstein01}.  The magnetic moments are thus plotted
vs the temperature of the simulation halved.  The staggered moment
(solid lines) decreases with increasing temperature and vanishes at a
N\'eel temperature which is about 4 times larger for the Tc
compound, in agreement with experiments. Also in agreement with
experiment, the Tc magnetic moment is smaller.  The plotted magnetic
moments are calculated from the magnetization on the orbitals $\psi_m$
by $\mu= \mu_B \sum_{m}
[n_{\uparrow}(\psi_m)-n_{\downarrow}(\psi_m)]$, where $n_\sigma(\psi)$
is the density of electrons on orbital $\psi$ with spin
$\sigma$. Because these orbitals have considerable weight on the
oxygen ions, the low-temperature moments exceed experimental values by
about 15\%. These Wannier orbitals are deliberately chosen to stress
that the suppression of the magnetic moment on Tc is not due to covalency
effects alone. The quantitatively correct low-temperature moments are
reported towards the end of this Letter.

\begin{figure}
 \begin{center}
   \includegraphics[width=0.95\columnwidth,keepaspectratio]{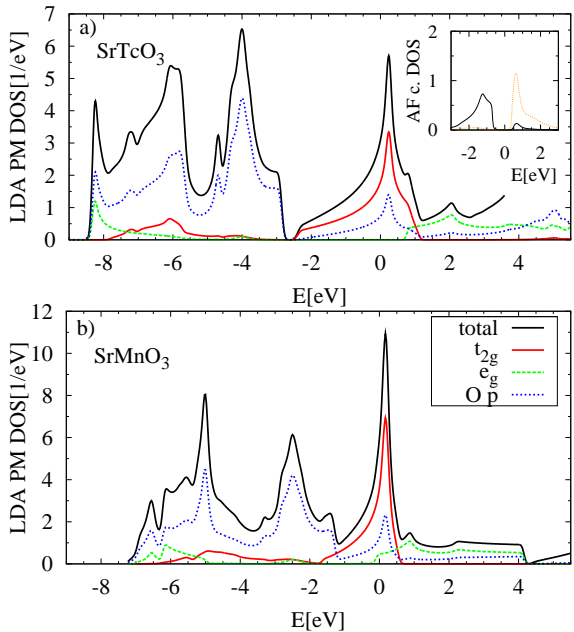}
   \end{center}
   \caption{\label{fig:dos} The total and orbitally resolved LDA
     paramagnetic (PM) DOS
     for (a) SrTcO$_3$ and (b) SrMnO$_3$. Inset: Predicted
     spin-resolved LDA+DMFT DOS.  }
\end{figure}

In Fig.~\ref{fig:dos}, we plot the total and orbitally resolved
paramagnetic LDA density of states (DOS). Both materials have three
\tg\ electrons, with the \tg\ manifold at the Fermi level, oxygen
states below, and \eg\ states above. The \tg\ bands constructed out of
Tc 4d orbitals are considerably broader with a bandwidth $W\approx
3.6$~eV, compared to 2.3\,eV found in 3d Mn.  Covalency, as measured
by comparing the oxygen weight in the \tg\ energy window is
significant in both compounds: The integrated oxygen DOS amounts to
about 20\% of the total integrated DOS for both cases. Once due to the
correlations the \eg\ states are pushed to higher energies, both
compounds have a half-filled \tg\ shell with a low lying $S=3/2$
multiplet.  Among 4d perovskites, in which \eg-\tg\ crystal field
splitting exceeds Hund's rule coupling, this largest possible spin is
realized only in the Tc compounds since other elements (Ru and Mo) are
not stable in the appropriate oxidation state.  In the inset, the
LDA+DMFT $t_{2g}$ DOS is plotted. The correlations open up a gap of
1eV, a bit smaller than 1.5eV found in hybrid-functional calculations
\cite{franchini11}. The gap in local spin-density approximation (LSDA)
is 0.3eV \cite{rodriguez11}. 

\begin{figure}
 \begin{center}
   \includegraphics[width=0.95\columnwidth,keepaspectratio]{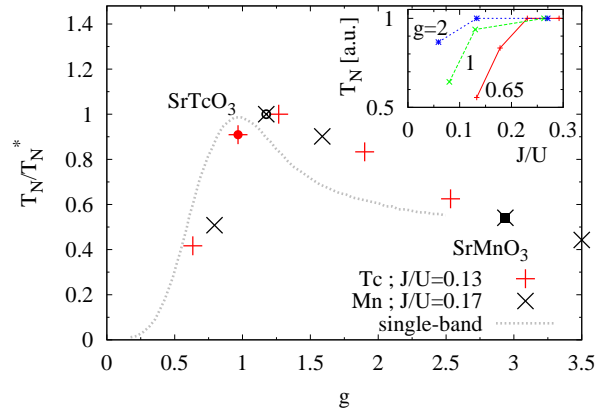}
   \end{center}
   \caption{\label{fig:trend} The N\'eel temperatures as a function of
     coupling $g\equiv{\cal U}/{\cal U}_c$, with ${\cal U}\equiv U+2J$
     and ${\cal U}_c=(U+2J)_c$ the value at which the MIT in the
     paramagnetic state occurs.  $T_N$ are normalized to the maximum
     N\'eel temperature $T_N^*$ found when varying ${\cal U}$.  The
     gray line is for the single-band Hubbard model (DMFT calculation
     of Ref.~\cite{rozenberg94}). Full circle and square
     indicate the data at physical value of interaction for the
     bandstructure of Tc and Mn compound, respectively. Empty circle
     denotes data for Mn compound, but artificially reduced
     interaction, see Fig.~1. Inset: The dependence of $T_N$ on
     $J/U$ at few $g$.}
 \end{figure}

While their large spin gives a hint about why Tc perovskites are
special among 4d oxides it does not tell why their 3d analogue
SrMnO$_3$ has a N\'eel temperature which is 4 times smaller. To
understand this, we first turn to qualitative considerations.  In the
localized picture, antiferromagnetism can be described in terms of the
Heisenberg model: $H=J_H \sum_{\langle i j \rangle} \mathbf{S}_i\cdot
\mathbf{S}_j$. The superexchange coupling $J_H$ between neighboring
spins $\mathbf{S}_i$ and $\mathbf{S}_j$ appears due to the
optimization of kinetic energy of the underlying itinerant model and
thus scales as $W^2/{\cal U}$ where $W$ is the bandwidth and ${\cal
  U}$ measures the strength of the Coulomb interaction.  On the other
hand, starting from the itinerant side the magnetism can be described
within a mean-field treatment of electronic interactions.  When
perfect nesting applies, this treatment yields an exponentially small
gap and transition temperature $\propto \exp (-W/{\cal U})$.  Hence, a
key observation is that robust magnetism occurs at intermediate values
of interaction ${\cal U} \approx W$ where the crossover from the itinerant
regime to the local regime takes place.  This has been established
firmly for the half-filled 3d Hubbard model, where the largest N\'eel
temperature $T_N$ is found very close to the metal-insulator
transition (MIT) in the paramagnetic state \cite{rozenberg94}
(cf. gray line in Fig.\ref{fig:trend}).

To relate this qualitative discussion to the realistic multiorbital
cases at hand, we calculated the properties of two compounds for a
range of fictitious interaction parameters. The $J/U$ values are fixed
to $0.17$ and $0.13$ for the Mn and Tc band structure, respectively.
A larger $J/U$ value for 3d compounds is used since the oxygen bands
are closer for 3d's and screen $U$ more efficiently than $J$.  We
denote the energy of the lowest atomic excitation of $S=3/2$ atomic
ground state by ${\cal U}=U+2J$.  ${\cal U}$ is the relevant quantity
for estimating where the system is located in the phase diagram with
respect to the MIT, and the Mott gap is ${\cal U}- c W$, where $c$ is
of the order of unity.  The MIT in the paramagnetic state occurs at a
critical value ${\cal U}={\cal U}_c$, where ${\cal U}_c=3.0\mathrm{eV}$
($U=2.4\mathrm{eV}$) for Tc and ${\cal U}_c=1.6\mathrm{eV}$
($U=1.2\mathrm{eV}$) for Mn, the larger ${\cal U}$ for Tc being due to
the larger bandwidth.  While the physical value of $U$ for Tc is
close to the value at which the transition takes place
\cite{rodriguez11,demedici11}, Mn is situated further to the
insulating side.  For non-half-filled $t_{2g}$'s, the critical ${\cal
  U}_c$ is increased by $5J$ \cite{[{See, e.g., }]demedici11a}, so
that most other $4d$ perovskites are paramagnetic metals.

In Fig~\ref{fig:trend}, we plot the N\'eel temperatures for Tc and Mn
bandstructures and several interaction strengths as a function of the
coupling $g\equiv {\cal U}/{\cal U}_c$.  The data are normalized with
respect to the maximum value $T_N=T_N^*$ found for a given
compound. Except at smallest $g$, where the details of the band
structure become more important, the data fall on the same curve and
can be described in terms of an universal function $T_N/W=f({\cal
  U}/W,J/U)$. Moreover, the dependence on $J/U$ (see the inset) is weak in
the regime relevant for the two compounds.  The maximum N\'eel
temperature is found at a bit larger coupling than in the
single-orbital Hubbard model but still close to the MIT.  Close to the
maximum, SrTcO$_3$ is found. On the localized side the N\'eel
temperature drops with increasing interactions, in accordance with
$J_H \propto W^2/{\cal U}$, because $T_N \propto J_H S$ in this
regime, and the Mn compound is found there.  As ${\cal U}$ is
diminished, the N\'eel temperature increases but more slowly than
$J_H$ does, because of the charge fluctuations which diminish the
effective value of the local spin.  $T_N$ reaches a maximum where
these fluctuations become substantial.  LDA+DMFT is the method of
choice to accurately describe materials such as SrTcO$_3$ which are in
the regime where neither the fully localized nor the fully itinerant
picture applies.

The maximal value of $T_N^* = W f(g^*,J/U)$ are thus reached for
interaction $g^* \approx 1$.  However, reducing the interaction
strength in SrMnO$_3$ to $g^*$ (Fig.~\ref{fig:moment}, open circles)
gives $T_N$ which still amounts only to about half of that found in
SrTcO$_3$. Most of the additional increase can be attributed to the
larger bandwidth of the Tc compound, setting the energy scale, as
recognized also in Refs.~\cite{rodriguez11, franchini11}.  The
occurrence of large $T_N$ of SrTcO$_3$ can thus be understood in two
steps, schematically depicted by arrows in Fig~\ref{fig:moment}.
First, $T_N$ is maximized for a given band structure as $g\to
g^*\approx 1$ (proximity to MIT), and, second, the change in the
band structure given mostly by the bandwidth accounts for the rest of
the enhancement.  These observations provide important clues for
engineering materials with robust magnetic properties.

It is instructive to consider the effect of an applied pressure, which
increases the overall bandwidth and also reduces the coupling $g$.
For SrMnO$_3$ both effects lead to an increase of $T_N$. In contrast,
for SrTcO$_3$ the two effects largely cancel each other.  To
substantiate this claim, we calculated $T_N$ for both compounds for
smaller volumes.  Lattice parameters $a$ were reduced by 1\%,
corresponding to about 5\,GPa \cite{zhou03}.  In SrMnO$_3$ we find a
relative increase of $T_N$ of 0.12, giving $-\alpha=d\log(T_N)/d \log
V \approx 3.7$, which is somewhat smaller than the experimental value
at small pressures \cite{zhou03}.  For the same compression, $T_N$ of
SrTcO$_3$ remains {\em constant}.  For larger compression with $a$
diminished by 2\%, $T_N/2$ {\it decreases} to 950K corresponding to
positive $\alpha$.  The qualitative conclusion that in SrTcO$_3$ the
value of $\alpha$ is small follows from the proximity of the compound
to the itinerant-to-localized transition and is thus robust against
uncertainties ($\sim20\%$) in the choice of interaction parameters.

The different position of the two materials on the ${\cal U}/{\cal
  U}_c$ diagram has also other consequences. The charge fluctuations
$\delta N^2 =\langle N^2 \rangle -\langle N\rangle^2$ are for
SrMnO$_3$ small $\delta N^2 <0.05$ as the material is situated well on
the localized side. Instead, for SrTcO$_3$, we find large $\delta
N^2=0.35$. Note that $\delta N^2=0.57$ for an atomic state
$|\psi\rangle$, which consists of 2,3, and 4 electron states with
equal probability 1/3.  The large value of the charge fluctuations for
SrTcO$_3$ is further evidence that a localized Heisenberg model
description \cite{franchini11} for these technetium compounds is
questionable.  These charge fluctuations explain also why the magnetic
moment of Tc is smaller. If maximal spin compatible with charge
fluctuations as in $|\psi\rangle$ is assumed, the local moment is
reduced to $7/3$\,$\mu_B$, which would lead to a suppression of the
moment by a factor of $7/9$. In reality, the charge fluctuations in Tc
are a bit smaller but still significant and crucial to explain the
smaller magnetic moment of Tc. The paramagnetic moment for the two
materials estimated from $\langle S_z^2 \rangle$ above $T_N$ is 2.7
for the Tc and 3.8 for Mn compound.  The paramagnetic magnetic moment
of Mn is close to the maximal $S=3/2$ moment $3.87$, while for Tc
charge fluctuations and related occurrence of states with smaller
spins suppress the moment.

\begin{table}
\begin{ruledtabular}
\begin{tabular}{cc   ccc   c}
case & LSDA & MT & on $\psi_m$ & on $\Psi_m$ & exp.  \\ 
\hline
SrTcO$_3$ & 1.3  & 1.4 & 2.5 & 2.2   & 2.1 \cite{rodriguez11} \\ 
SrMnO$_3$ & 2.3  & 1.9 & 3.0 & 2.6   & 2.6 \cite{takeda74} \\
\end{tabular}
\end{ruledtabular}
\caption{(2nd-4th columns) Low-temperature staggered magnetic moments
  in units of $\mu_B$, calculated within LDA+DMFT as obtained within
  muffin-tin (MT) spheres, on the extended Wannier function $\psi_m$,
  constructed without oxygen states, and on the localized Wannier
  function $\Psi_m$ constructed by including all the oxygen states
  (see the text).  These values are compared to LSDA (1st column) and
  experiment (last column).
  \label{table2}}  
\end{table}

Finally, we turn to the delicate question of how to quantitatively
determine the magnetic moment as measured by experiment. Magnetic
moments obtained by different means are reported in
Table~\ref{table2}. Within LDA the moments are usually determined from
the magnetization found in the muffin-tin spheres. This approach gives
reasonable values for Mn compound, which only slightly underestimate
experimental values, but fails badly for the Tc compound. This occurs
because the larger 4d orbitals extend outside the muffin-tin
spheres. Similar discrepancies with experiments are seen also if we
calculate the magnetic moments within muffin-tin spheres in our
approach. A less biased method is to determine the moment from the
magnetization on the Wannier orbitals. If the small energy window
orbitals $\psi_m$ are used, these orbitals contain significant weight
also on oxygens (due to covalency) and thus overestimate the
experimental values.  To obtain quantitative agreement with
experiment, one must construct well-localized Wannier orbitals from an
energy window including also oxygen bands. The magnetic moment found
on these orbitals $\Psi_m$ agrees with experiment within our
precision. These results demonstrate unambiguously that (i) in
SrMnO$_3$ the $\sim 15\%$ suppression of the magnetic moment from maximal
3$\mu_B$ is due to the covalency, and (ii) in SrTcO$_3$ similar covalency
effects are indeed seen but that (iii) most of the suppression of
magnetic moment actually occurs due to the larger charge fluctuations
in this much more itinerant compound.

\paragraph{Conclusion}
In summary, we have calculated the properties of SrTcO$_3$ in the
magnetic state with a combination of band-structure methods with the
dynamical mean-field theory. We have shown that SrTcO$_3$ lies close
to an itinerant-to-localized transition, which explains its high
N\'eel temperature. Other 4d oxides do not have a half-filled $t_{2g}$
shell and are thus further away from the transition, or they have
distorted crystal structures, which reduces the bandwidth and
suppresses the kinetic energy gain related to the formation of
magnetic states. We have shown that the charge fluctuations in
SrTcO$_3$ are large, which explain the smallness of the measured
magnetic moment and implies that a purely localized (Heisenberg)
description is not applicable. To put these results into perspective,
we have analyzed the isostructural and isoelectronic 3d- SrMnO$_3$
within the same framework.  This compound has a smaller bandwidth and
lies further to the localized side which explains its much smaller
N\'eel temperature and larger magnetic moment, despite similar
covalency.  The physical differences between the two materials imply
very different pressure dependence of $T_N$, which is predicted to be
weak for SrTcO$_3$.  Taken together, these results help understand the
occurrence of robust magnetism and may provide clues for designing
magnetic materials.

\acknowledgements{ We thank L. de'Medici, L. Pourovskii, and
  L. Vaugier for useful discussions.  Support was provided by the
  Partner University Fund (PUF) and the Swiss National Foundation
  MaNEP program. M.A. acknowledges financial support by the Austrian
  Science Fund (FWF), Project No. F4103. J.M. acknowledges support
  from the Slovenian Research Agency under Contract No. P1-0044.}

\end{document}